\documentclass[runningheads]{cl2emult}

\usepackage{makeidx}  
\usepackage{graphicx} 
\usepackage{subeqnar} 
\usepackage{multicol} 
\usepackage{cropmark} 
\usepackage{eso}      
\makeindex            

\begin{document}
\title*{Clusters of Galaxies in the 2dF Galaxy Redshift Survey\footnote[1]{{\bf The 2dFGRS Team:} Carlton M.\ Baugh (Durham), Joss Bland-Hawthorn,
Terry Bridges, Russell Cannon (AAO), Shaun Cole (Durham), Matthew Colless
(ANU), Chris Collins (LJMU), Nicholas Cross (St. Andrews), Gavin Dalton
(Oxford), Kathryn Deeley (UNSW), Simon P.\ Driver (St. Andrews), George Efstathiou (Cambridge), Richard S.\ Ellis (CalTech), Carlos S.\ Frenk
(Durham), Karl Glazebrook (JHU), Carole Jackson (ANU), Ofer Lahav
(Cambridge), Ian Lewis (AAO), Stuart Lumsden (Leeds), Steve Maddox
(Nottingham), Darren Madgwick, Stephen Moody (Cambridge), Peder Norberg
(Durham), John A.\ Peacock (Edinburgh), Will Percival (Nottingham),
Bruce A.\ Peterson, Ian Price (ANU), Will Sutherland, Helen Tadros
(Oxford), Keith Taylor (CalTech)}}
\toctitle{Clusters of Galaxies in 2dFGRS}
%
%
\titlerunning{2dFGRS Galaxy Clusters}
%
\author{Roberto De Propris\inst{1}
\and Warrick J. Couch\inst{1}
\and the 2dFGRS Team}
\authorrunning{Roberto De Propris et al.}
%
%
\institute{Department of Astrophysics and Optics\\
           University of New South Wales\\
           Sydney, Australia, 2052}

\maketitle              

\begin{abstract}

The 2dF Galaxy Redshift Survey has obtained 135,000 redshifts for
galaxies in two broad strips. Here we present the first results of
a 3-dimensional search for galaxy clusters based on known 2-dimensional
compilations. We derive new redshifts and velocity dispersions for
clusters, assess the level of contamination in the sample, analyze
the accuracy of photometric redshift estimates and study the space
distribution of clusters.

\end{abstract}

\section{Introduction}

It has long been known that the distribution of galaxies is not
homogeneous over scales of at least 200 $h^{-1}$ Mpc; the first
redshift surveys revealed an intricate pattern of filaments, voids and
walls (e.g., Da Costa 1999) and showed that samples of the order
of 10$^5$ galaxies were needed to reach the scale of homogeneity and
derive cosmologically useful quantities from analysis of the
3-dimensional distribution of galaxies. The 2dF Galaxy Redshift Survey
(e.g., Colless 1999) is the first of the new generation of surveys to
be able to obtain such samples with reasonable efficiency. Among papers
being published by the 2dFGRS Team it is worth mentioning: the $b$ band
type-selected luminosity function (Folkes et al. 1999), the $K$-band
luminosity function from 2MASS photometry (Cole et al. 2000), the
bivariate brightness distribution (Cross et al.  2000), an accurate
estimate of the $\beta$ parameter (Peacock et al. 2000) and Principal
Component Analysis of galaxy populations (Madgwick et al., this volume
and 2000).

Clusters of galaxies are the largest bound structures in the observable
universe and the only ones that can be observed and identified to
cosmologically significant redshifts. The mass distribution of
clusters yields limits to the cosmological density parameter; for
instance, the standard Cold Dark Matter (CDM) model, normalized to
the COBE data, yields cluster densities in excess of observations
by an order of magnitude and this provided the first hint of an
open universe model. 

However, most studies of clusters are still based on photographic
catalogs, such as those of Abell and collaborators (1958;
1989), or the APM (Dalton et al. 1992) or Edinburgh-Durham (Collins
et al. 1995). Once the 2dF survey is complete, it will be possible 
to determine a catalog of groups of galaxies on scales extending
from compact groups to giant clusters, using 3-dimensional selection
algorithms. While this is not easily done at this stage, it is
possible to consider clusters selected from 2-dimensional catalogs
in 3-dimensional space and assess their reality and level of contamination,
derive their space density and study the properties of their members.
In turn, this will allow us to 'define' a cluster for later 3-dimensional 
searches, matching the properties of known template objects.

\section{Selection of Clusters}

We have used known cluster catalogs and matched the cluster centroid
and search radii to the 2dF redshift catalogs. The procedure we followed
is described in detail in De Propris et al. (2000) and is a modified
form of the 'gapping' algorithm used by Zabludoff et al. (1993). In
summary, we identify isolated peaks in redshift space and compute
redshifts and velocity dispersions using an iterative process. We
also carry out the same analysis for all significant peaks found in
the cluster line-of-sight, as defined by its associated search radius.
Figure 1 shows cone plots and redshift histograms for three representative
objects.

\begin{figure}
\centering
{\includegraphics[width=1.0\textwidth]{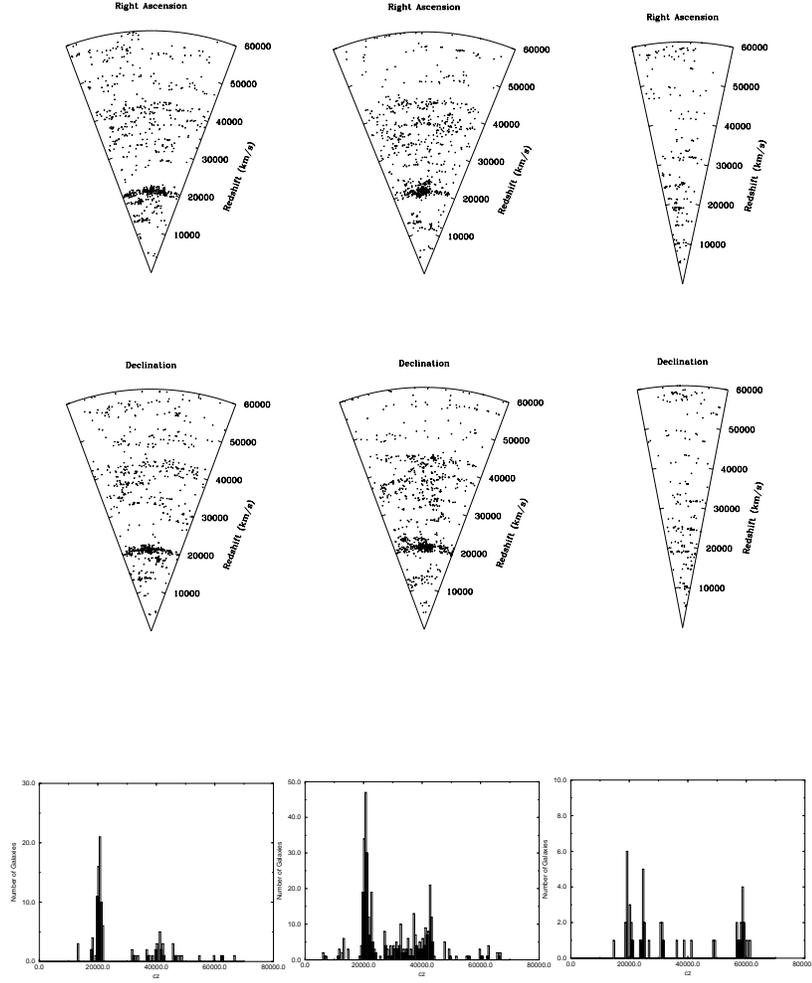}}
\caption[]{Cone diagrams (6 Mpc opening angle) and redshift histograms
(over the Abell radius) for S0333 (first from left), Abell 3094 (middle)
and Abell 3824 (right) as examples of (a) a well-defined isolated object
(S0333); (b) a cluster with significant structure in its line of sight
(A3094) and (c) a cluster resolved in numerous groups (A3824)}
\label{eps1}
\end{figure}

A summary of cluster identifications is presented in Table 1: we also
cross-identify objects present in more than one catalog. The total
number of unique objects in our study is 233, of which 123 are present
on only the Abell catalogues, 24 in the APM and 86 in the EDCC. 

\begin{table}
\centering
\caption{Summary of Cluster Identifications}
\renewcommand{\arraystretch}{1.4}
\setlength\tabcolsep{5pt}
\begin{tabular}{llll}
\hline\noalign{\smallskip}
Catalog & N[Clusters(Abell,APM,EDCC)] & N(Redshifts) & N($\sigma$) \\
\noalign{\smallskip}
\hline
\noalign{\smallskip}
Abell N & 51(--,7,13) & 30 & 17 \\
Abell S & 159(--,22,63) & 107 & 53 \\
APM & 54(29,--,25) & 43 & 22 \\
EDCC & 169(76,25,--) & 115 & 67 \\ \hline
Total & 433 & 294 & 159 \\  \hline

\end{tabular}
\label{Tab1a}
\end{table}

About 1/3 of all clusters are not yet identified in 2dFGRS; in most
cases this appears to be due to the fact that clusters are either poor
(richness class 0) or very distant (with m$_{10}$ -- where this is the
magnitude of the 10$^{th}$ brightest galaxy, used as a redshift
indicator -- values indicating $z > 0.12$). In some instances the
clusters lie in a low completeness region and redshifts may become
available at a later stage.

Important to any quantitative analysis based on the clusters found here
is the need to identify volume-limited subsamples and correct for 
incompleteness due to our window function and selection efficiency.
Two routes have generally been adopted: selection of candidates based
on 'cuts' in $m_{10}$, which defines a roughly volume limited sample
and with a richness limits that makes the sample reasonably complete,
and a pure redshift selection, now possible from 2dF data.  The former
technique has been the one most generally used and it is therefore
appropriate to consider its accuracy and limitations.

Figure 2 plots estimated vs. 2dF redshifts for all three catalogs being
considered (we separate the Abell and Abell et al. catalogs as they
are selected differently). We see that, whereas there is a broad 
relation between real and estimated redshift, there are numerous
objects where the estimators fail and all catalogs saturate at some
level, where $m_{10}$ approaches the plate limit.

\begin{figure}
\centering
\rotatebox{270}
{\includegraphics[width=.65\textwidth]{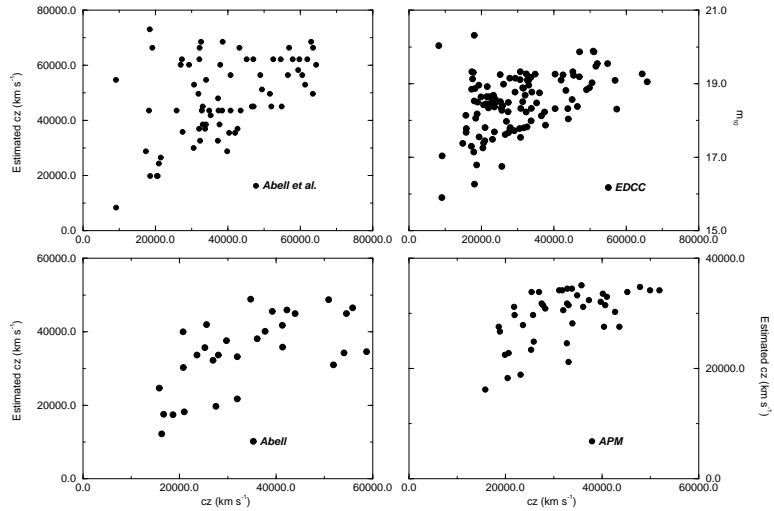}}
\caption[]{Comparison of estimated and real redshifts}
\label{eps1}
\end{figure}

The broad relation apparent in Figure 2 can be used to define an
estimated $cz$ such that, given the spread in the relation, one can
select an approximately volume limited sample which is also reasonably
complete. We can use this to choose cuts in estimated redshift space
and define samples for studies of contamination, where we define
contamination as the presence of significant foreground and/or
background structure in the line of sight. We define this structure
to be significant if our 'gapping' procedure described above yields 
a redshift or velocity dispersion for any of the secondary peaks.

By this definition we find that the Abell and EDCC samples are
contaminated at approximately the 15\% level; the APM catalog 
suffers at only the 5\% level, but this is likely a factor of
the larger richness cut and smaller search radius used.

Figure 3 plots the space density of clusters in all three catalogs
as a function of the cluster redshift; we can use these plots
to define a real redshift where the samples are complete; since
the space density of clusters is believed to be constant, at least
over the volume sampled, any apparent decline in density may be
attributed to the onset of incompleteness. We plot the X-ray selected
sample of RASS1 (De Grandi et al. 1999) for comparison and to show
the apparently constant density of clusters. Our data appear to be
complete to about $z=0.11$ (not coincidentally, the peak in the redshift
distribution for the whole survey); this is then chosen as the
redshift limit for our estimated redshift cut (and for our pure
$cz$ sample). Note that in both the EDCC and Abell survey there is
a relative lack of clusters at $z=0.05$ that may be related to the
'hole' claimed by Zucca et al. (1997), although it is most likely
an artefact of the small volume sampled so far.

\begin{figure}
\centering
\rotatebox{270}
{\includegraphics[width=.65\textwidth]{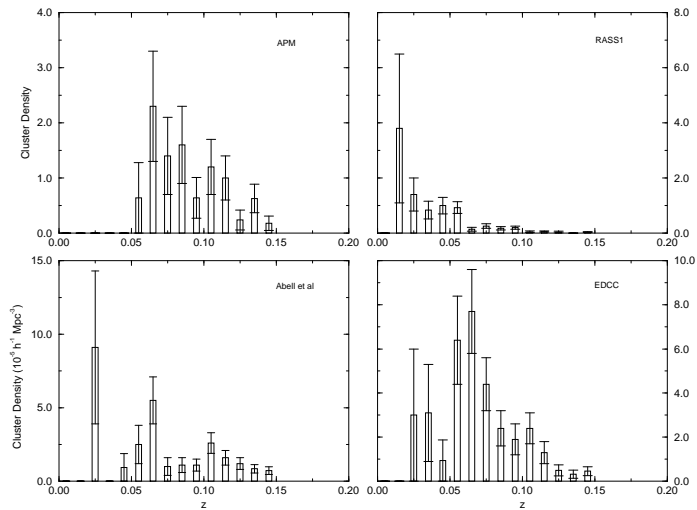}}
\caption[]{Space Density of clusters in all catalogs}
\label{eps1}
\end{figure}

\section{The Space Density of Abell Clusters and the Distribution of $\sigma$}

We select an estimated 'cut' such that objects with real $cz < 33000$
km/s are included. However, Figure 2 shows that many such objects are
actually excluded from the sample. We correct for this by using clusters
whose real $cz < 33000$ km/s and calculating how many of these have
estimated $cz > 33000$ km/s. We obtain a total space density for all
Abell clusters of $26.1 \pm 3.5 \pm 7.6$ (where the second error is
the error due to our completeness correction and the units are
$10^{-6}$ $h^3$ Mpc${-3}$) and $4.9 \pm 1.5 \pm 1.8$ for $R > 1$ clusters.

We also choose a sample of clusters with $cz < 33000$ km/s using only
those objects with measured redshifts. While this is certainly incomplete
it provides a reliable lower limit to the space density of clusters.
We obtain, for all Abell clusters, $19.4 \pm 2.7$ and, for $R > 1$
objects, $7.8 \pm 1.8$. This latter result is similar to the values
determined by Zabludoff et al. and Mazure et al. (1996).

The distribution of velocity dispersions provides some constraint
on models of structure formation, via the shape of the power
spectrum of fluctuations. Cluster masses, in particular, provide
limits in small scales and help in normalizing Cosmic Microwave
Background results. Whereas estimating cluster masses is extremely
difficult, the distribution of velocity dispersions may be used
as a substitute, especially at the high end, which is most sensitive
to cosmology.

We plot our data in Figure 4 together with previous compilations.
Although these comparisons should be taken with some caution,
especially at the low end, where our sample includes low richness
objects, they should be fair at the high end, where we observe
reasonable agreement. 

\begin{figure}
\centering
\rotatebox{270}
{\includegraphics[width=.65\textwidth]{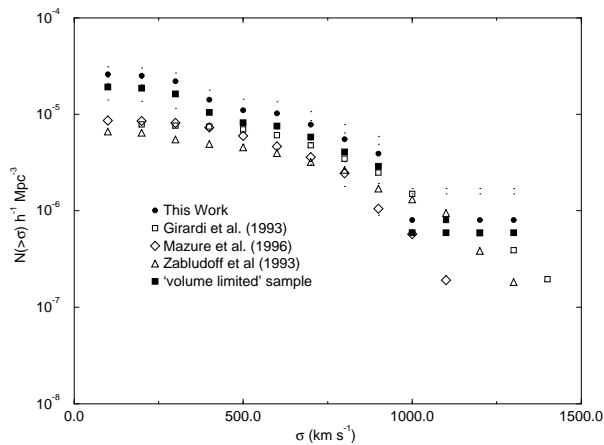}}
\caption[]{Distribution of cluster velocity dispersions for our
data (filled symbols) and previous work as indicated in legend}
\label{eps1}
\end{figure}

The most robust result we can derive is the relative lack of clusters
of high velocity dispersion; indeed, since contamination will increase
the derived velocity dispersion, we feel that we can determine a 
significant upper limit to the space density of N($\sigma > 1000$)
clusters of $ < 2.5$ (in the same units as above). Our $cz$ sample
also allows us to derive a weak lower limit of 0.85. These can be
compared with theoretical models by Borgani et al. (1998): we find
that our data suffice to rule out Standard CDM models, $\Lambda$CDM
models with high $\Omega_M$ and $\tau$CDM cosmologies, while allowing
Cold and Hot dark matter models, open CDM and low $\Omega_M$ $\Lambda$CDM. 
Our data are therefore in favor of low values of $\Omega_M$, which
would indeed bring cluster data in better agreement with the COBE
and CMB results.

\vskip 1cm

\clearpage
\addcontentsline{toc}{section}{Index}
\flushbottom
\printindex

\end{document}